\documentstyle{elsart}

\begin{document}
\date{26 November 1998}
\begin{frontmatter}
\title{Discrete  random  walk  models  for \\
  symmetric  L\'evy - Feller   diffusion  processes}

\author[Berlin]{Rudolf Gorenflo},
\author[CINECA]{Gianni De Fabritiis} and
\author[Bologna]{Francesco Mainardi\thanksref{mail1}}
\address[Berlin]{Department of Mathematics  and Computer Science,
   Free University of Berlin, D-14195 Berlin, Germany}
\address[CINECA]{CINECA, Supercomputing Center,
  I-40033 Casalecchio di  Reno, Bologna, Italy}
\address[Bologna]{Dipartimento di Fisica, Universit\`a di Bologna
 and INFN, Sezione di Bologna,
   Via Irnerio 46, I-40126 Bologna, Italy}
\thanks[mail1]{corresponding author, mainardi@bo.infn.it}

\begin{abstract}
We propose a variety of  models of random walk,
discrete in space and time, suitable for simulating
stable random variables of arbitrary index $\alpha$ ($0<\alpha \le 2$),
in the symmetric case.
We show that by properly scaled transition
to vanishing space and time steps
 our random walk models
converge to the corresponding continuous Markovian stochastic processes,
that we refer to as L\'evy-Feller diffusion processes.
\end{abstract}


\end{frontmatter}

\def\eg{{\it e.g.}\ } \def\ie{{\it i.e.}\ }
\def\sg{\hbox{sign}\,}
\def\sgn{\hbox{sign}\,}
\def\sign{\hbox{sign}\,}
\def\e{\hbox{e}}
\def\exp{\hbox{exp}}
\def\ds{\displaystyle}
\def\dis{\displaystyle}
\def\q{\quad}	 \def\qq{\qquad}
\def\lan{\langle}\def\ran{\rangle}
\def\l{\left} \def\r{\right}
\def\lra{\Longleftrightarrow}
\def\arg{\hbox{\rm arg}}
\def\d{\partial}
 \def\dr{\partial r}  \def\dt{\partial t}
\def\dx{\partial x}   \def\dy{\partial y}  \def\dz{\partial z}
\def\rec#1{{1\over{#1}}}
\def\log{\hbox{\rm log}\,}
\def\erf{\hbox{\rm erf}\,}     \def\erfc{\hbox{\rm erfc}\,}
\def\F{\hbox{F}\,}
\def\NN{\hbox{\bf N}}
\def\RR{\hbox{\bf R}}
\def\CC{\hbox{\bf C}}
\def\ZZ{\hbox{\bf Z}}
\def\II{\hbox{\bf I}}

\section{Introduction}
By   a	{L\'evy-Feller} diffusion process we mean a
Markovian process governed by a   stable probability
density function ($pdf$) evolving in time,
$g_\alpha (x,t;\theta)\,, $ whose spatial Fourier transform
(the characteristic function) reads
$$\hat g_\alpha(\kappa,t;\theta) =\int_{-\infty}^{+\infty} \!\!
  \e^{\,\ds  i\kappa x}\, g_\alpha(x,t;\theta)\, dx=
\exp\, \l(-t|\kappa|^{\alpha}\,
\e^{\, \ds i(\sg \kappa)\theta \pi/2}\r)\,,
  \eqno(1.1) $$
where $ x\,,\,\kappa \in \RR\,,\, t>0\,.$
The two relevant parameters,
$\alpha \,, $ called the {\it index of stability},
and $\theta $ (related to the asymmetry), improperly referred to
as the {\it skewness},	are real numbers
subject to the conditions, see \eg  \cite{Feller71},
$$ 0 <\alpha \le 2\,; \qq
   |\theta| \le \cases{
   \alpha \,, &if $\q 0<\alpha <1\,, $\cr
   2-\alpha \,, &if $\q 1\le \alpha \le 2\,. $\cr}\eqno(1.2)$$
By introducing
the similarity variable  $x\,t^{-1/\alpha }\,, $
we can write
$  g_\alpha(x,t;\theta)=
 t^{-{1/\alpha}}\,p_\alpha(x\,t^{-1/\alpha};\theta)\,, $
where $p_\alpha (x;\theta )$ is the stable $pdf$ at $t=1\,. $
The specific form of the characteristic function (1.1) allows us
to recognize
$g_\alpha(x,t;\theta)$ as the Green function
(fundamental solution) of the Cauchy problem
$$
{\partial \over \partial t}\, u(x,t) = D^\alpha_\theta \, \l[u(x,t)\r]\,,
\q u(x,0) =\delta (x)\,,
\q x\in {\RR}\,,\q t>0\,,
\eqno(1.3)
$$
where $D^\alpha_\theta\,$
is the pseudo-differential operator
 with  symbol
$$  \widehat D_\theta ^\alpha =
    -|\kappa|^{\ds \alpha} \, \e^{\,\ds i(\sg \kappa)\theta \pi/2}\,.
\eqno(1.4) $$
Let us recall that
a generic pseudo-differential operator $A$,
acting with respect to the variable $x \in \RR\,,$
is defined through its Fourier representation, namely
  $    \int_{-\infty}^{+\infty}
  \e ^{\, i\kappa x} \,  A \,[ \phi (x)] \, dx =
 \hat A(\kappa )\, \hat \phi (\kappa )\,,  $
  where $\phi(x)$ denotes a sufficiently well-behaved function in
 $\RR\,,$  and $\hat A(\kappa)\,$
is referred to as  symbol of $A\,,$
   given as
 $ \hat A (\kappa ) = \l( A\, \e^{\, -i\kappa x}\r)\,
  \e^{\, +i\kappa x}\,. $

With the names of L\'evy and Feller we have intended
to honour both Paul L\'evy, 
\cite{Levy37}, who first introduced the
class of stable distributions,
see  \cite{Levy24}, \cite{Levy25}, \cite{Levy37},
and William Feller \cite{Feller52},
who first investigated the semi\-groups generated  by a
pseudo-differential  equation of type (1.3-4).
For $\alpha =2$ and $\alpha =1$ (with $\theta =0$)
we recover the standard Gaussian and Cauchy $pdf's$
$$g_2(x,t;0) =
{1\over 2\sqrt{\pi}}\, t^{-1/2}\, \exp\left( -{x^2\over 4t}\right)\,,
\q
 g_1(x,t;0)={1\over \pi}\, {t\over x^2 + t^2}\,. \eqno(1.5)
$$
In physics, the first recognition that the (symmetric) L\'evy
distribution could be characterized via  a pseudo-differential
operator of type (1.3-4) was made explicitly by West \& Seshadri
\cite{West82}.
Recently, Gorenflo \& Mainardi \cite{GM98a}, 
\cite{GM99a},	 \cite{GM99b},
 have revised  Feller's original
arguments by interpreting (1.3)
as a space-fractional diffusion equation (of order $\alpha $
and "skewness" $\theta $)  and have
provided a   variety of related random walk models, discrete
in space and time,
 which by properly scaled transition
to vanishing space and time steps
 converge to the corresponding continuous
 L\'evy-Feller  processes.
In other words the   discrete probability distributions
generated by the random walk models have been proved to belong to
the domain of attraction of the corresponding stable distribution.

Here, limiting ourselves to the symmetric case ($\theta =0$),
we  present the main features  of
the random walk models by Gorenflo \& Mainardi,
and we display preliminary
results of a few numerical case studies, which
can be of some interest  in  econophysics.
The  L\'evy statistics in
modelling fluctuations of economical and financial variables,
formerly used by Mandelbrot in the early sixties,  is still
nowadays  adopted with success, see \eg
\cite{Mandelbrot97},
\cite{MS97},  \cite{BP97}.


\section{Outline of the general theory}
      Let $Y$ be an integer-valued  random variable
and let   the random variables
$Y_1, Y_2, Y_3, \dots  $ be
i.i.d. (= {\it independent identically distributed}),
all having their  probability distribution common with $Y$ .
We define  a spatial-temporal grid
$\{(x_j\,,\, t_n)   \, \vert\, j\in \ZZ\,, \; n \in \NN_0 \} $
by
 $\, x_j = x_j(h) = j\,h\,,$ $\,t_n = t_n(\tau ) = n\, \tau \,, $
where $h>0$  and $\tau >0\,. $
Then we consider the sequence of random variables
$$S_n=hY_1+hY_2+\dots+hY_n,\q n\in{\NN}\,,
\eqno(2.1)
$$
with (for convenience) $S_0 =0\,,$
and interpret it as follows.
A particle, sitting in $x=x_0 = 0$
at time $t=t_0 =0$ finds itself at a later instant
$t=t_n$ in point
$x = S_n$ which is
an integer multiple of $h\,. $
We recognize the
$p_k= P(Y = k)$  (for $k \in \ZZ$)
as "transition probabilities":
$p_k$ is the probability
of a particle jumping from a point $x_j= S_n$
to a point $x_{j+k} = S_{n+1}$	as time proceeds
from $t_n$ to $t_{n+1}\,. $
All $p_k$ are non-negative, and their sum equals 1.

The probability $y_j(t_n)$
of sojourn of our particle in point
$x_j$ at instant $t_n\,$ obeys the
{\it transition law}
$$y_j(t_{n+1})=\sum_{k= -\infty}^{+\infty}
  p_k\, y_{j-k}(t_n)\,,\q    y_j(0) = \delta _{j\,0}\,,
\qq j\in {\ZZ}\,,\q n\in {\NN}_0\,.
\eqno(2.2)
$$
which has the form of a discrete convolution.
Hence,
introducing {\it generating functions}
 $$\tilde p(z) = \sum_{j= -\infty}^{+\infty}  p_j z^j,\q
 \tilde y_n(z) = \sum_{j= -\infty}^{+\infty}  y_j(t_n) z^j,
\eqno(2.3)
$$
we obtain
 $$\tilde {y}_n (z) =
     \tilde {y}_0 (z) \cdot  [\tilde  p(z)]^n
  = [\tilde  p(z)]^n \,, \; n\in \NN_0 \,. \eqno(2.4)$$

The power series in (2.3) and (2.4) are
absolutely and uniformly convergent on
$|z|=1\, $ and assume the value 1 at $z=1\,. $
Putting
 $ z=\e^{i\kappa\, h}\,,\, \kappa\in {\RR}\,,$
and observing
$ z^j=\e^{i\kappa\, jh}= \e^{i\kappa \, x_j}\,, $
we recognize
$\hat  p(\kappa;h)= \tilde p(\e^{i\kappa h})$ and
$ \hat y (\kappa,t_n;h)= \tilde y_n(\e^{i\kappa h})$
as {\it characteristic functions}
of the random variables $h\,Y$ and $S_n\,, $ respectively.

Our aim is to approximate the L\'evy-Feller diffusion process,
which is governed  by the evolution equation (1.3),
arbitrarily well.
To this purpose we introduce a strictly monotonic {\it scaling relation}
$ \tau	= \sigma (h) \to 0$ as $h\to 0\,. $
We will fix $t>0$ and let  $h$
(and likewise $\tau $) go to zero over such values that always
$ n = t/\tau = t/\sigma (h)$  is a positive  integer.
Then  we have the equivalences
 $$ n \to \infty
   \, \Longleftrightarrow \,
 h \to 0  \, \Longleftrightarrow \,  \tau \to 0\,,  $$
and  $h$ depends on $\tau $,
finally on $n$, so that $h = h(n)\,. $
Replacing $h$ by $h(n)$ in (2.1) we obtain
a sequence of random variables
$X_n$
with characteristic functions
$ \hat y(\kappa ,t_n;h) = [\tilde p(\e^{i\kappa h})]^n\,$
(note that now $t_n =t$ is fixed).
Invoking Theorem 3.6.1 of Lukacs  \cite{Lukacs60},
what remains  to be shown is that
$\,\hat y(\kappa ,t;h) \to \exp (-t|\kappa |^\alpha )\,$
as $h \to 0\,,$
the characteristic function of the corresponding symmetric
L\'evy-Feller process.
For this it  suffices  that, for
fixed $\kappa  \ne 0\,, $
$$ \log \l[\hat y(\kappa ,t;h)\r] \equiv
 {t\over \sigma (h)}\, \log \big[ \tilde{p}(\e^{i\kappa h})\big]
  \to -t\, |\kappa |^\alpha \,, \q \hbox{\rm{as}}\q h\to 0\,.
\eqno(2.5)$$
Our random walk  can  be interpreted as a "difference scheme"
to approximate the evolution equation( 1.3),
if we write (2.2) in the
equivalent form, observing the scaling relation,
$$
{y_j(t_{n+1})- y_j(t_n)\over \tau } =
 \rec{\sigma (h)} \,
\big[ (p_0-1)\, y_j(t_n) + \sum_{k\ne 0} p_k \, y_{j-k}(t_n) \big]\,,
\eqno(2.6)$$
In fact, whereas  the  L.H.S. is the explicit discrete approximation
to the	first-order time  derivative ${\d \over \dt} u(x,t)\,, $
the R.H.S  can be considered as a particular
 discrete  approximation to
the space pseudo-differential term
$D_0 ^\alpha \,[ u(x,t)]\,, $ provided we mean
$ y_j(t_n)   = \int_{x_j-h/2}^{x_j+h/2} \! u (x,t_n) \, dx
  \approx  h\, u (x_j,t_n )\,$
with	$ y_j(0) = \delta _{j\,0}\,. $

\section{The  random walk models}
From the previous Section we have learnt that,
in order to construct discrete random walk models which are
convergent (in distribution) to the (symmetric) stable $pdf'$s,
the clue points are:
1) to guess
a suitable {\it generating  function}  $\tilde p(z)\,, $
whose coefficients of its power series expansion
provide the transition probabilities,
2) to determine the corresponding {\it scaling relation}
 $\tau = \sigma (h)$ which ensures the required convergence.

In the classical case of the Gaussian distribution ($\alpha =2$)
the matter is easily treated if we remember that the corresponding
density is the fundamental solution of the standard diffusion equation,
which  is known to be well approximated
via the finite-difference equation
$$
{y_j(t_{n+1}) -y_j(t_n)\over \tau} =  {
 y_{j+1}(t_n)- 2 y_{j}(t_n) + y_{j-1}(t_n)\over h^2}\,,
\q y_j(0) = \delta_ {j \, 0}\,.
   \eqno(3.1)$$
In this case, introducing the scaling parameter
 $ \mu = {\tau / h^2}\,,$ so
 $\tau =\sigma (h) = \mu \, h^2\,,$
 the transition probabilities turn out to be
$$p_0 = 1 - 2 {\mu}\,, \q
 p_{\pm 1} =  {\mu}\,, \q
 p_{\pm k} = 0\,, \q k = 2,3 \dots\,,. \eqno(3.2)$$
subject to the condition
$   0< {\mu }\le 1/2\,.$
Thus the generating function is
 $$
\widetilde p(z) = 1+\mu [ z-2 +  {z}^{-1}]  \,. \eqno(3.3)$$
The proof of the convergence to the Gaussian is simple
since one easily finds
$  [t/(\mu h^2)]\, \log [\widetilde p(\e^{i\kappa h}]
     \to -t\kappa^2 $ as $h\to 0\,.$
The scheme  (3.2) means that for approximation of the standard
Gaussian process the corresponding random walk model
exhibits  only jumps of one step to the right or
one to the left or jumps of width zero.
For the stable non Gaussian processes we expect to find a
non-polynomial generating function with infinitely many
transition coefficients which imply the  occurrence of
arbitrarily large jumps. It is common practice to
refer to the corresponding random walks  as to
{\it  L\'evy flights}.

In the following, limiting ourselves to the symmetric cases ($\theta =0$),
we shall resume  the main features of three different random walk models,
referred to as (RW1), (RW2), (RW3),
  of which Gorenflo
and Mainardi have proved the  convergence to the corresponding
continuous processes.
 For  each model we  give
the generating function $\widetilde p(z)$
with the transition probabilities $p_k$ and
the scaling relation $\tau  = \sigma (h)$,
 referring  to the original
papers for details.
From the analysis of the classical Gaussian case
 we find it natural to first introduce
the scaling parameter $\mu = \tau /h^\alpha $ with
$0<\alpha \le 2$ but, as we shall show later,
  this will not necessarily imply
   $ \sigma (h) = \mu \,h^\alpha\,,  $
with $\mu $ constant for fixed $\alpha \,. $

The  model (RW1) has been introduced and discussed in
 \cite{GM98a}, 
starting from the identification
of the operator $D_0^\alpha  $
in the framework of fractional calculus, see \cite{Feller52},
and then applying, in the authors' original approach,
the  Gr\"unwald-Letnikov  discretized scheme.
For this model we need to keep distinct the two cases
(a) $0<\alpha <1$ and (b) $1<\alpha \le 2\,, $
being the case $\alpha =1$   excluded in this treatment.
The   generating function is
$$\tilde p (z)=  \cases{
   1-{\ds{\mu\over 2\cos(\alpha\pi/2)}}\,
       [(1-z)^\alpha+(1-z^{-1})^\alpha]\,, &
      $0<\alpha < 1\,,$ \cr
   1-{\ds{\mu\over 2\cos(\alpha\pi/2)}}\,
     [z^{-1}(1-z)^\alpha+ z(1-z^{-1})^\alpha]\,, &
     $1<\alpha \le 2\,.$ \cr }
\eqno(3.4)
$$
For both cases the scaling relation   is confirmed to be
$ \tau = \sigma (h) = \mu \, h^\alpha  \,, $
but the parameter $\mu $ is subject to different
restrictions to ensure that $ 0 \le p_0 <1\,. $

In the case (a) we have $\,0< \mu\le \cos(\alpha\pi/2)\,$
and
$$
\cases{
 p_0=1- {\ds{ \mu\over \cos(\alpha\pi/2)}}\,, & \cr
 p_{\pm k}=(-1)^{k+1}\, {\ds{\mu\over 2 \cos(\alpha\pi/2)}}\,
   {\ds{\alpha \choose k}} \,, &
  for  $\q k=1,2,3,\dots\,.$ \cr}
\eqno(3.5a)$$
In the case (b) we have
$ \,0 < \mu  \le {|\cos(\alpha\pi/2)|/\alpha}\,$   and
 $$\cases{
 p_0 =1- {\ds{\mu \alpha \over |\cos(\alpha\pi/2)|}}\,, \q
 p_{\pm 1}   ={\ds{\mu\over 2|\cos(\alpha\pi/2)|}} &
  $\l[{\ds{\alpha \choose 2}} + 1\r]\,,$  \cr
 p_{\pm k}= (-1)^{k+1}\, {\ds{\mu\over 2|\cos(\alpha\pi/2)|}}\,
   {\ds{\alpha \choose k+1}} \,,
   &  for $\; k=2,3,4,\dots$\cr} \eqno(3.5b)  $$
We note that, whereas the classical
 Gaussian random walk (3.2) is
prompt\-ly recovered from (3.5b) for $\alpha =2$,
the random walk for the Cauchy process	($\alpha =1$)
 cannot be obtained, neither directly nor
by a passage to the limit $\alpha \to 1\,. $
Indeed, in both the limits $\alpha \to 1^-$ and $\alpha \to 1^+$
the permissible range of the scaling factor $\mu $ is vanishing.
In numerical practice the consequence will be that if $\alpha $ is
near 1 the convergence is slow: for good approximation
we will need a very small step-time $\tau $ with respect
to the step-length $h\,. $

Differently from (RW1) we shall show how
the two other  random walk models,
(RW2) and  (RW3), that we are going to briefly discuss,
exhibit a smooth scaling law in the range $ 0<\alpha <2$,
so the case $\alpha =1$ is no longer singular.
However, "there  is no free lunch". We have to pay for the
good behaviour at $\alpha =1$ with bad behaviour at $\alpha =2$
in a sense to be seen later.

The  model (RW2) is obtained by imposing for the transition
probabilities an expression in terms of binomial coefficients,
as suggested from (3.5b),
which is required to be regular in the limit  as $\alpha  \to 1\,,
$ namely
   $$
 p_0=1- 2\lambda \,, \q
 p_{\pm k}= (-1)^{k+1} {\lambda \over \alpha -1}\,
  {\alpha \choose k+1} \,,
  \q \hbox{for } k=1,2,3,\dots\,,
\eqno(3.6) $$
where $\lambda\,,  $  subject to the condition $0<\lambda \le 1/2$
is to be determined.
The model has been extensively discussed in   \cite{GM99b}
whereas the particular case $\alpha =1$ (related to the Cauchy process)
 has been formerly presented  in \cite{GM99a}.
 To guarantee the convergence  for all $\alpha $
($0<\alpha \le 2$) to the corresponding continuous process,
the asymptotics as $h \to 0$ still requires a scaling
relation  of the kind
$\tau  = \mu \, h^\alpha $   (with  $\mu $ constant),
namely
$$\tau	= \sigma (h) =
  \cases{
  {\ds{2 \lambda \cos (\alpha \pi/2) \over  1-\alpha}}\, h^\alpha
	 &if $\q 0<\alpha <2\,,\; \alpha \ne 1\,,  $ \cr
  \lambda \, \pi \, h  &if $\q	\alpha = 1\,,  $ \cr
   \lambda\, h^2     &if $\q \alpha =2 \,. $\cr} \eqno(3.7)$$
We recognize that for $0  <\alpha <2$
the model (RW2) exhibits a smooth  scaling law
but  a discontinuity is present at $\alpha =2$ as
can be seen by taking the limit as $\alpha \to 2^-\,. $
This  allows us to recover for $\alpha =2$
the Gaussian model (3.2)-(3.3),
but it shows that  in numerical practice
when $\alpha$ is near 2
the convergence is expected to be slow.
For $\,0<\alpha< 2\,$  we obtain the generating functions
and the transition probabilities as follow.
Putting $ \, \rho (z) =  [(1-z)^{\alpha -1} -1]\,$
for $\alpha  \ne 1\,, $
and introducing the scaling parameter  $\mu\,,	$
we have
$$  \widetilde p(z) =\cases{
    1- {\ds{\mu \over 2 \cos (\alpha \pi/2)}}
  \,\l[ (1-z^{-1})\,\rho (z) + (1-z)\,\rho (z^{-1})\r],
    & $ \alpha \ne 1\,,  $ \cr
  1 - {\ds {\mu \over \pi}}\,
[(1-z^{-1})\,\log (1-z)+(1-z)\,\log (1-z^{-1})],
  & $ \alpha =1\,.  $ \cr}
 \eqno(3.8)$$
For $0< \alpha <2 \,,\; \alpha \ne 1\,$ we have
  $\,0< \mu \le  \cos(\alpha\pi/2)/(1-\alpha)\,$  and
 $$  \cases{
 p_0 =1- {\ds{\mu (1-\alpha) \over \cos(\alpha\pi/2)}}\,,& \cr
 p_{\pm k}= (-1)^{k}\, {\ds{\mu\over 2\cos(\alpha\pi/2)}}\,
   {\ds{\alpha \choose k+1}} \,, &
  for  $\q k=1,2,3,\dots\,,$ \cr}    \eqno(3.9) $$
for $\,\alpha =1\,$ we have   $\,0< \mu \le \pi /2\,$ and
$$ p_0 = 1 - {2\mu \over \pi} \,,\q
  p_{\pm k} =  {\mu \over \pi} \, {1\over k(k+1)} \,,
  \q \hbox{for}\q k=1, 2,3,4,\dots\,.  \eqno(3.10) $$

In both  models (RW1), (RW2) the generating function is
expressed in terms of elementary   functions and the transition
coefficients, for $0<\alpha <2\,, $
exhibit an  asymptotic behaviour
as $ |k|  \to \infty$ consistent  with that of the power-law
tails of the stable densities as $|x| \to \infty\,,$ see \eg
\cite{Feller71}. Indeed we obtain
 $$ p_{k} \sim	 \mu  \,
    \Gamma(\alpha +1) \,
 {\sin (\pi\, \alpha/2)\over \pi} \, |k|^{-(\alpha+1)}
  \q \hbox{as}\q  |k| \to \infty\,,\q 0<\alpha <2\,;
\eqno(3.11) $$
$$ p_\alpha (x;0) \sim
     \Gamma (\alpha +1)\, {\sin(\pi\alpha/2)\over \pi}\,
  |x|^{-(\alpha+1)} \q \hbox{as}\q
  |x| \to\infty\,,  \q 0<\alpha <2
\,.\eqno(3.12)	$$

The asymptotic behaviour of the stable densities is the starting
point for the random walk model (RW3) in that, following
Gillis \& Weiss \cite{Weiss70}, we require that all $p_k$
for $k\ne 0$ are proportional to $|k| ^{-(\alpha +1)}\,. $
However, the simplicity of the starting point leads
to  difficulties for treating this model
since the  generating function is no longer  elementary,
a fact that makes the convergence proof a really hard affair.
We need to recall the following special functions
$$  \zeta (\beta ) = \sum _{k = 1}^\infty  {k^{-\beta }}\,, \;
  \beta  > 1\,,\q
 \Phi (z,\beta) = \sum _{k=1}^\infty {z^k\over k^{\beta }}\,, \q
 |z|<1\,, \q \beta \in \RR\,, \eqno(3.13)$$
respectively known as the  Riemann {\it zeta} function
and the  {\it polylogarithmic} function.
This model has been extensively discussed in   \cite{GM99b}
(see also  \cite{Weiss70}),
where the following generating function is derived
$$ \tilde p(z)
 = 1- 2 \lambda \, \zeta (\alpha +1)
 + \lambda \, \l[\Phi(z, \alpha +1) + \Phi(z^{-1}, \alpha
+1)\r]\,.\eqno(3.14)$$
Here $\beta= \alpha +1 >1 $
so $ \Phi (z,\beta)$ is by its power series also defined on the
periphery $|z|=1$ of the unit circle.
Then we get a  pure power-law  random walk with
$$
p_0 = 1-   2 \lambda \, \zeta (\alpha +1)\,, \q
p_{k} = \lambda \, |k|^{-(\alpha +1)}\,, \q
   \hbox{for} \q k \ne 0\,,
\eqno(3.15)$$
where	 $\lambda\,,  $  subject to the condition
$ 0 < \lambda  \le 1/[2\,\zeta (\alpha +1)]\,, $
is to be determined.
To ensure the convergence to the corresponding
continuous process
  the following scaling relation must hold
$$\tau = \sigma (h)
 =  \cases{
{\ds{\lambda\,\pi \over \Gamma(\alpha +1)\,\sin(\alpha\pi/2)}}\, h^\alpha
	   &if $\q 0<\alpha <2\,, $ \cr
 \lambda \,h^2\, |\log h|   &if $\q  \alpha = 2\,.  $ \cr
    } \eqno(3.16)$$
It is interesting to note that in the case $\alpha =2$ the
classical random walk  model (3.2)-(3.3) is no longer
recovered, since now arbitrarily large jumps occur
(with a probability decaying as $k^{-3}$)
as in the L\'evy flights. Nevertheless, through the scaling relation
(3.16), this random walk converges to the continuous Gaussian process.
Thus this  model
gives us the opportunity to verify
that a random walk with infinite variance steps may (slowly) converge
to the Gaussian, in agreement with a general theorem
on the domain of attraction of the normal law,
see \eg \cite{Gnedenko54}.

\section{Numerical results}
In general the random walk models are not
only valuable from the conceptual point of view for visualizing
what the  diffusion means but
also for numerical calculations, either as Monte Carlo simulation of
particle paths in a diffusion process or as discrete imitation of the
process in form of redistribution (from one time level to the next) of
clumps of an extensive quantity (across the spatial grid points).
Our  models can be used in at least three different ways:
$(a)$ as {\it finite difference schemes} for approximate calculation
     of symmetric stable densities;
$(b)$ for producing {\it sample paths} of individual particles
      performing the random walk;
$(c)$ for producing {\it histograms}
     of the approximate realization of the densities $g_\alpha $ by
    simulating many individual paths with the same number of time steps
    and making statistics of the final positions of the particles.


For  numerical simulations of stable random
variables different algorithms have been provided
by a number of specialists,
including Chambers {\it et al} \cite{Chambers76},
 Bartels \cite{Bartels78},
 Mantegna \cite{Mantegna94},
 Janicki \&  Weron \cite{Weron94},
 Samorodnitsky \& Taqqu \cite{Taqqu94}.
Our present approach for treating
L\'evy statistics
has been carried out independently from all the
above  references but  uniquely based
on the random walk models presented here,
so, as far as we know, our results would be in the great part original.

Having preliminarily checked a sufficient level
of  accuracy for our
finite dif\-fer\-ence schemes
with the existing tables of stable densities,  here
we present some  results  on the simulation
of the sample paths and histograms corresponding to some typical
values of the index of stability,
namely $\alpha = 1, 1.5, 2\,. $
In practice, in our numerical studies
there is required truncation in two ways.
It is impossible to simulate  all infinitely many discrete probabilities,
so the size of possible jumps must be limited to a maximal
possible jump length. The other truncation is required if
a priori one wants a definite region of space to be considered
in which the walk takes place. Then, particles leaving
this space have been ignored.
 \begin{figure}
 \vspace{7truecm}
 \includegraphics{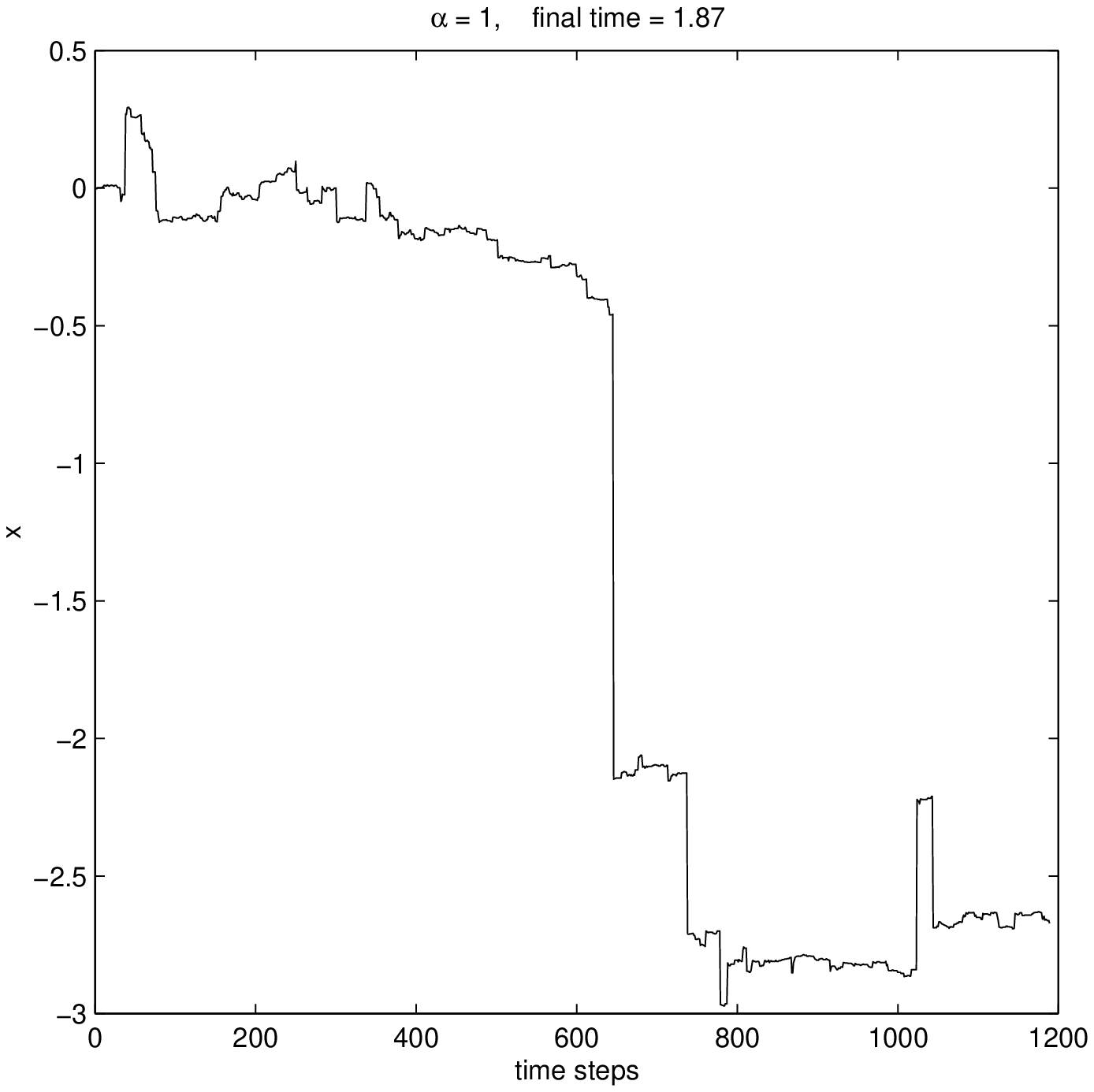}
 \includegraphics{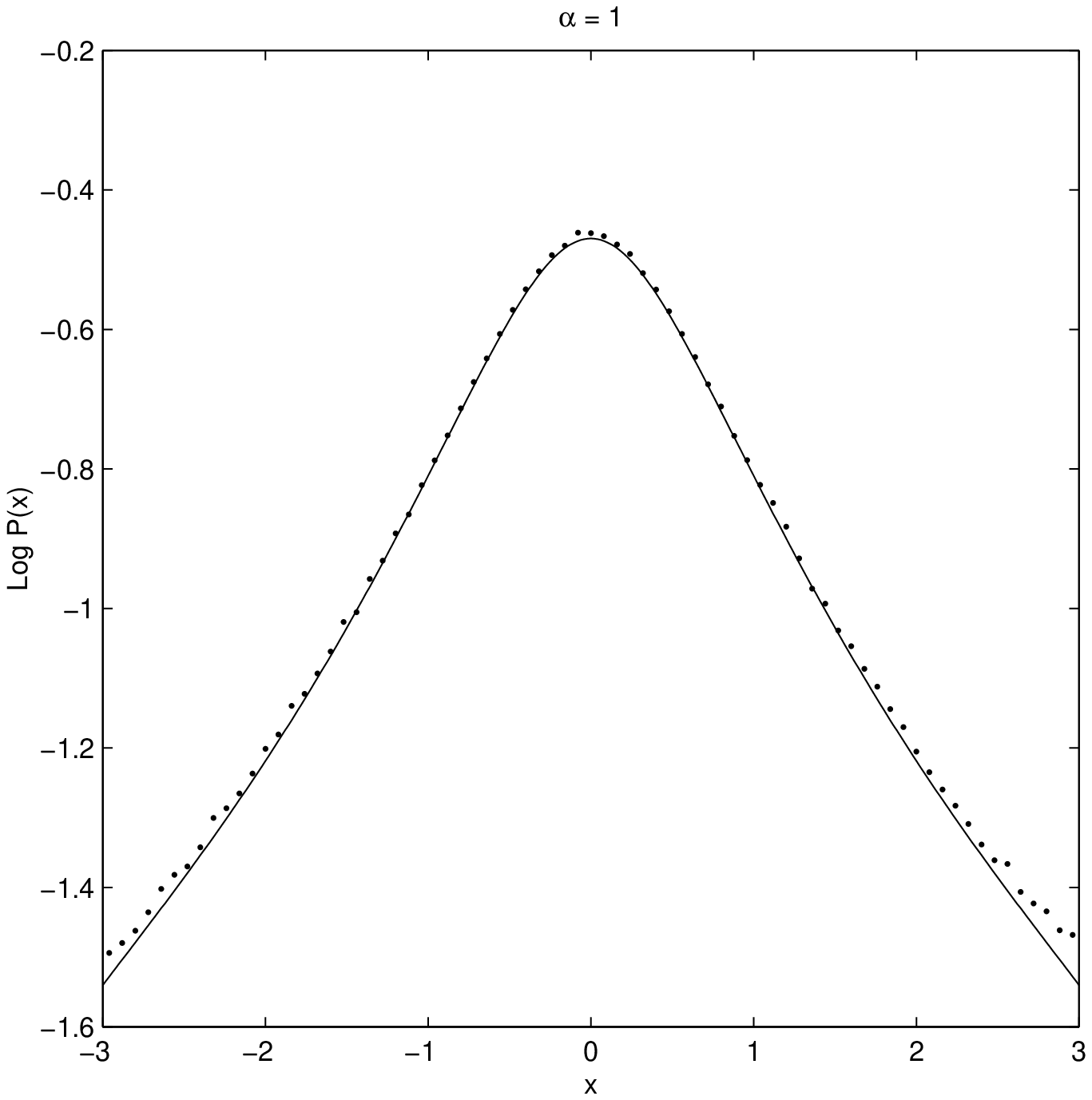}
 \caption
{Sample path (left) and histogram (right) for $\alpha =1\,.$ (Cauchy)}
 \bigskip
 \end{figure}
\begin{figure}
\vspace{7truecm}
\includegraphics{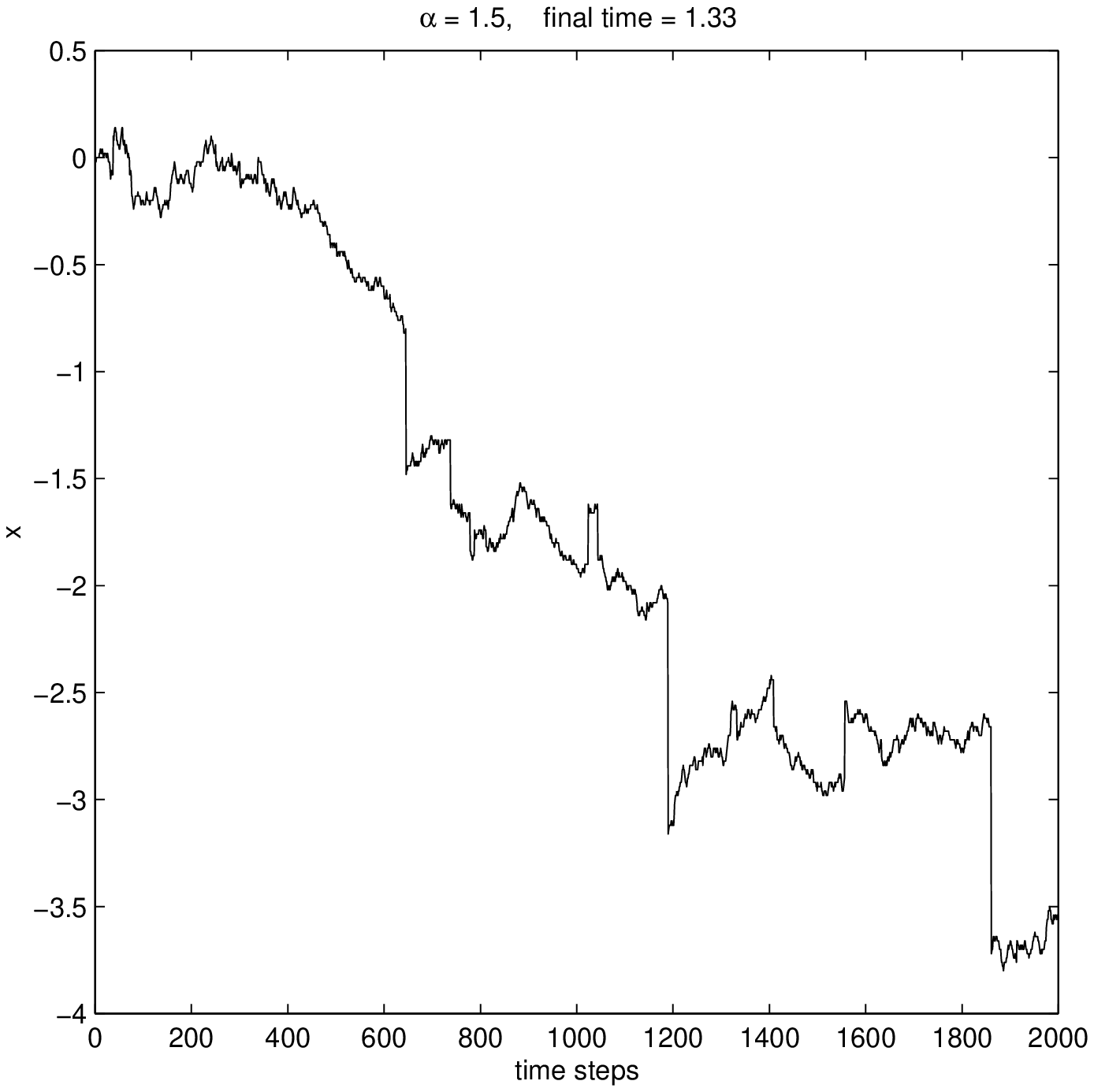}
\includegraphics{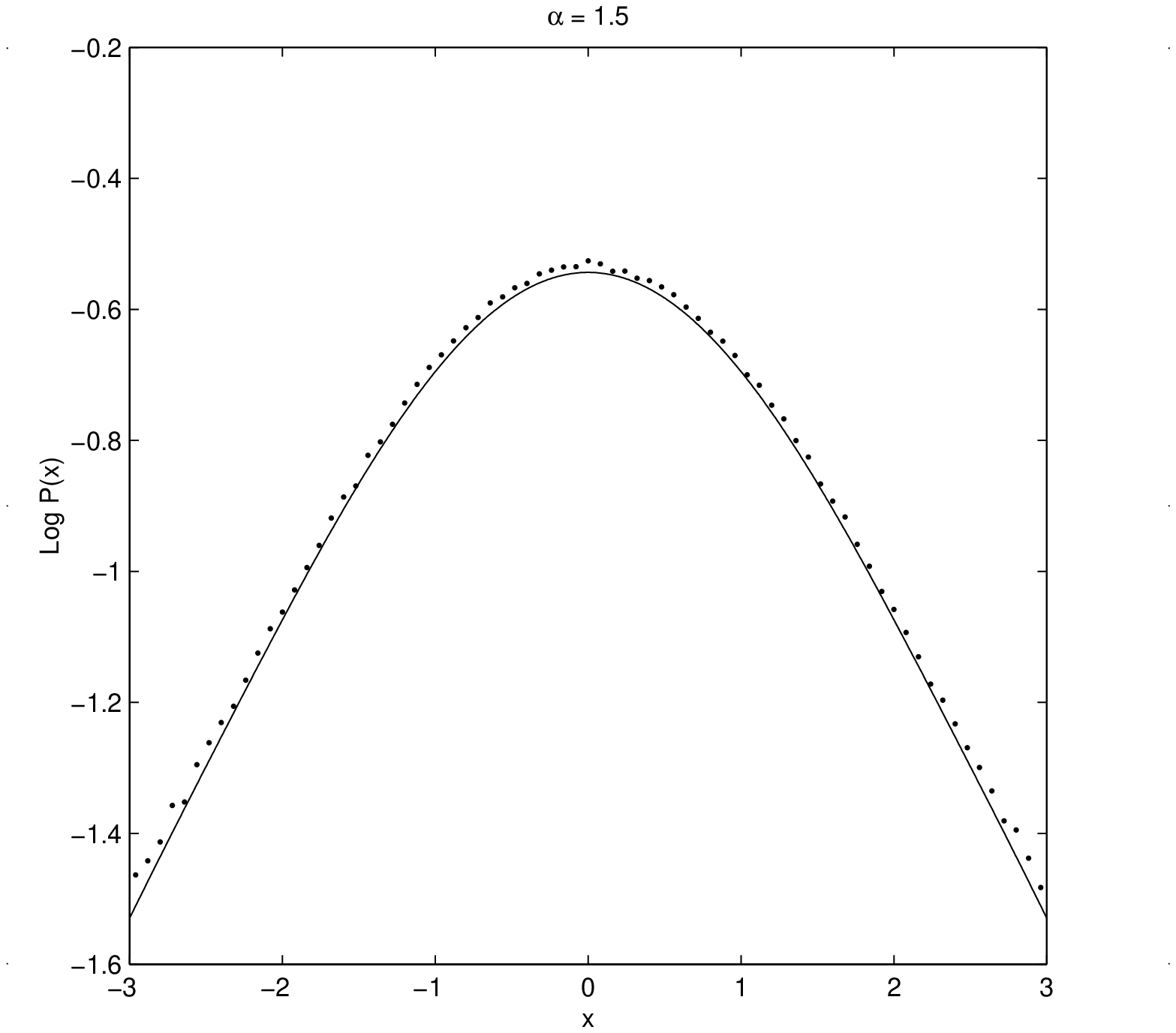}
\caption
{Sample path (left) and histogram (right) for $\alpha =1.5\,. $}
\bigskip
\end{figure}
Our simulations, based
on one million of realizations, have been carried out
in the interval $|x| \le 4\,.  $
All the histograms refer to stable densities at $t=1\,$
for $|x| \le 3\,, $ the space interval being reduced
to avoid  the border effects.
The sample paths  are plotted against the time steps,
up to 1200 for $\alpha =1$ and up to 2000 for $\alpha =1.5\,,2\,,$
so they refer to different final times, namely $t= 1.87, 1.33, 0.8\,,$
respectively.
The transition probabilities
have been chosen from our random walk models as follows:
$\alpha =1$ from (RW2), see (3.10),
with scaling parameter $\mu = \pi /4\,$;
$\alpha =1.5\,, $ from (RW1), see (3.5b),
with $ \mu  = (2/3) \cos (3\pi /4)\,;  $
for $\alpha =2\,, $ from the standard model, see (3.2),
with $\mu =1/4\,. $
The  cases   $\alpha =1$ (Cauchy process)
and $\alpha =2$ (normal process) have been considered
for a possible comparison with	the standard  and accurate
algorithms existing in the literature, whereas $\alpha =1.5$
has been chosen in view of possible applications
in econophysics where usually the index of stability
ranges from 1.4 to 1.7,
see \eg \cite{Mandelbrot97}, \cite{MS97}.

 \begin{figure}
 \vspace {7 truecm}
 \includegraphics{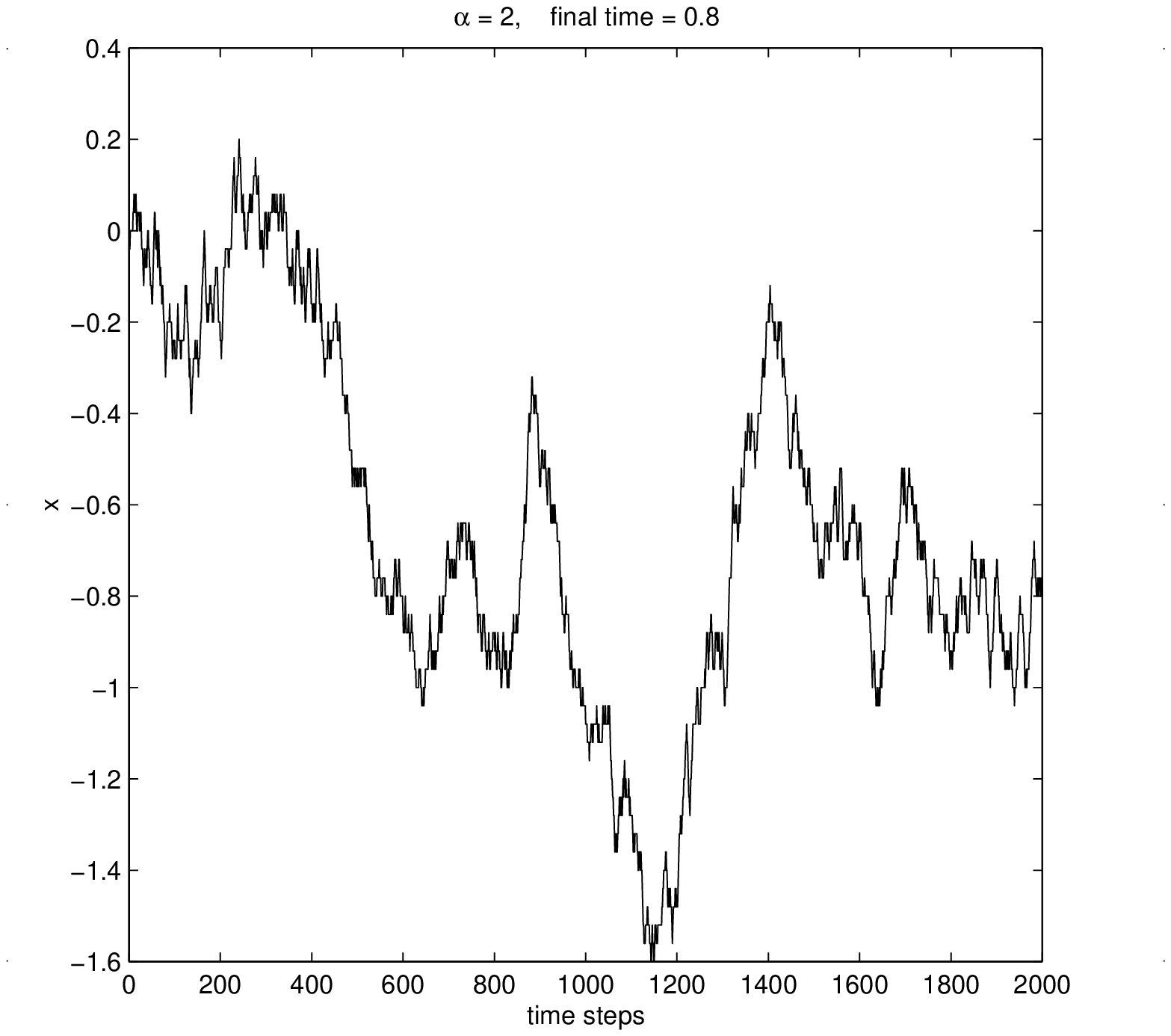}
 \includegraphics{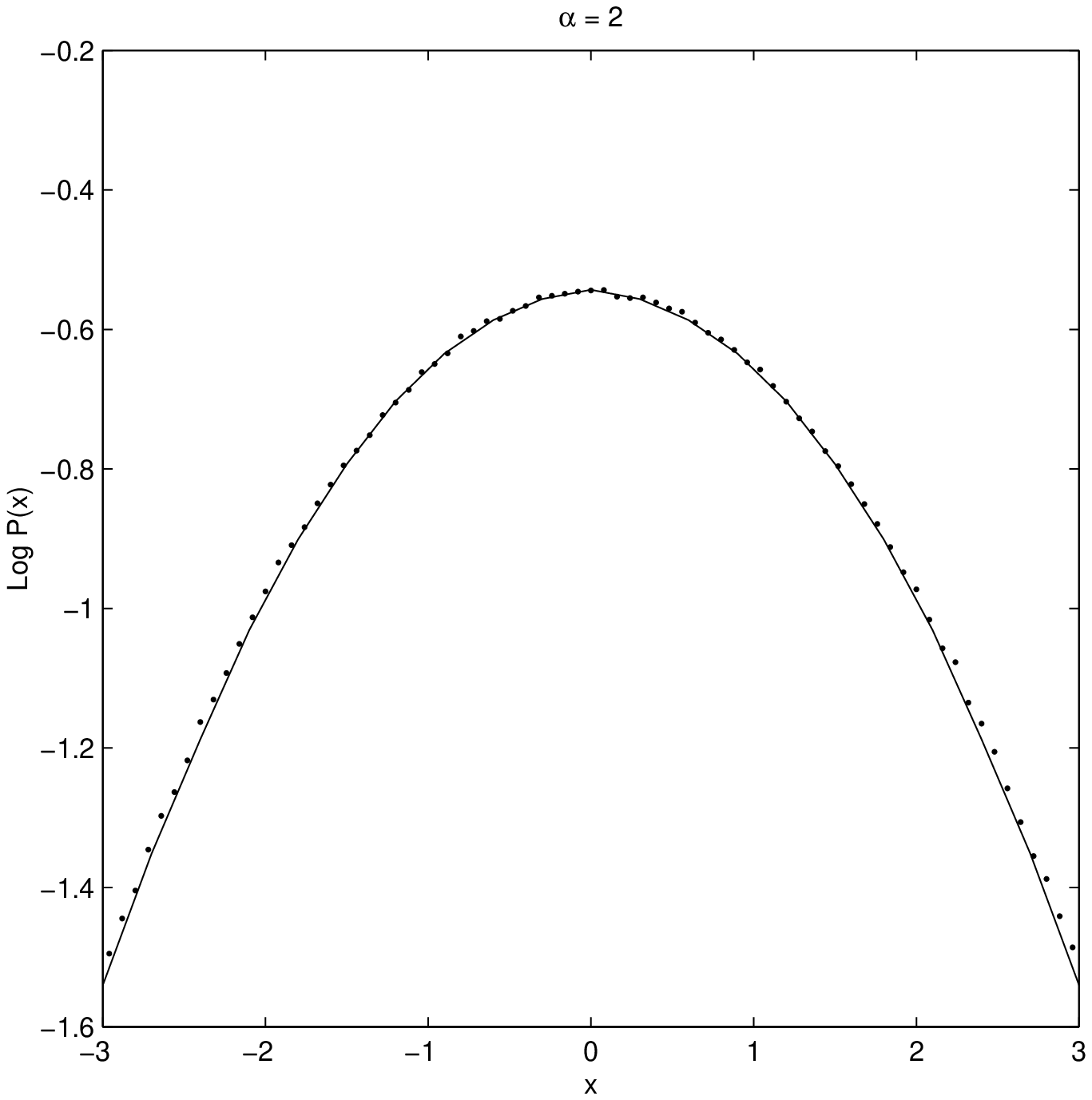}
 \caption
 {Sample path (left) and histogram (right) for $\alpha =2\,.$ (Gauss)}
  \bigskip
 \end{figure}

\section{Conclusions}

For the simulation of Markovian processes characterized
by symmetric L\'evy probability densities evolving in time
we have presented three different random walk models,
discrete in space and time, by giving their respective
transition probabilities. We have indicated how, by use of
generating functions, convergence for properly scaled transition
to vanishing steps of space and time can be analyzed, and we have
hinted at peculiarities occurring at the particular
stability indices $\alpha =1$ (the Cauchy process) and
$\alpha =2$ (the Gauss process). We have displayed preliminary
results of a few numerical case studies concerning sample
paths and histograms to check the efficiency
of our	algorithms.
From the sample paths
one can recognize the "wild" character of the L\'evy flights
with respect to the "tame" character
of the	Brownian motion.

We  expect that  our arguments
can be relevant  in different fields of physics including the emerging one
of  econophysics, where
stable distributions are becoming more common.
In  statistical physics the stable distributions
play a key role in the (wonderful) world of random
walks constructed by  the late Montroll and  continued through his
school, see \eg \cite{Montroll79}, \cite{Montroll84},
\cite{Klafter96}.
Also   Tsallis and his associates
have recognized the key role of stable distributions with respect to a
generalized theory of thermostatics, see \eg \cite{Tsallis97},
and references therein.
Here, we have (only briefly) pointed out the
relation between L\'evy statistics  and  space-fractional diffusion
equations. However, stable distributions turn out to
be related also with  time-fractional diffusion equations,
see \eg \cite{Budapest97}. Furthermore,
this topic is relevant for fractal phenomena, where
differential equations of fractional
order are usually adopted to describe their evolution,
see \eg    \cite{Carpinteri97}, \cite{Zaslavsky98},
\cite{Hilfer98}, and reference herein.

\begin{ack}

We gratefully acknowledge fruitful discussions with Rosario Mantegna.
This work was supported in part by the Italian CNR and INFN,
and by the Research Commission of Free University of Berlin.

\end{ack}


\end{document}